\documentstyle[nato,psfrag,graphicx,numreferences,amssymb,amsmath]{crckapb} 

\renewcommand{\Im}{\text{Im}}

\newcommand{\bs}{\boldsymbol}
\newcommand{\eps}{\varepsilon}
\newcommand{\beps}{\boldsymbol\varepsilon}
\newcommand{\bbeps}{\boldsymbol{\bar\varepsilon}}

\newcommand{\lpar}{l\,{\parallel}\,l}  
\newcommand{\lperp}{l\,{\perp}\,l}     
\newcommand{\hpar}{h\,{\parallel}\,h}	
\newcommand{\hperp}{h\,{\perp}\,h}      

\newcommand{\ket}[1]{|#1\rangle}
\newcommand{\bra}[1]{\langle#1|}
\newcommand{\mv}[1]{\left\langle#1\right\rangle}
\newcommand{\eq}[1]{\begin{equation} #1 \end{equation}}
\newcommand{\eqlab}[2]{\begin{equation} #1
	\label{#2.eq}\end{equation}}
\newcommand{\refig}[1]{Fig.~\textup{\ref{#1.fig}}}
\newcommand{\refeq}[1]{\textup{(\ref{#1.eq})}}

\newcommand{\Je}{J_{\mathrm e}}
\newcommand{\me}{m_{\mathrm e}}
\newcommand{\Jg}{J}
\newcommand{\sfT}{{\mathsf{T}}}
\newcommand{\sixj}[6]{\left\{\begin{array}{ccc}
		#1	& #2	& #3\\
		#4	& #5	& #6
		\end{array}\right\}}
\newlength{\argwidth}
\newlength{\vertexheight}
\newcommand{\lstep}{%
		\settowidth{\argwidth}{${\otimes}$}%
		\makebox[0.75\argwidth]{%
		\raisebox{\vertexheight}[2em][1.5em]{%
	 	\begin{picture}(0,25)(0,0)
			\put(0,0){\makebox(0,0){$\otimes$}}
			\put(0,25){\makebox(0,0){$\otimes$}}
			\qbezier[8](0,5)(0,12.5)(0,20)
		\end{picture}}}}

\newcommand{\avgintens}{%
	\raisebox{\vertexheight}[2em][1.5em]{%
        \begin{picture}(25,25)(0,0)
		\linethickness{0.255em}
		\put(0,0){\line(1,0){25}}
		\put(0,25){\line(1,0){25}}
	\end{picture}}}

\newlength{\intensboxheight}
\newcommand{\intensbox}[1]{%
	\setlength{\intensboxheight}{1.3125em}
	\setlength{\fboxsep}{0.097em}
	\framebox{%
          \raisebox{0pt}[\intensboxheight][-\vertexheight]{$\ #1\ $}}}

\newcommand{\crossed}{%
	\raisebox{\vertexheight}[2em][1.5em]{%
        \begin{picture}(40,25)(0,0)
 	     	\qbezier[12](9,4)(20,12.5)(31,21)
      		\qbezier[12](9,21)(20,12.5)(31,4)
		\put(5,0){\makebox(0,0){$\otimes$}}
		\put(5,25){\makebox(0,0){$\otimes$}}
		\put(35,0){\makebox(0,0){$\otimes$}}
		\put(35,25){\makebox(0,0){$\otimes$}}
		\linethickness{0.255em}
		\put(8.5,0){\line(1,0){22.5}}
		\put(8.5,25){\line(1,0){22.5}}
    	\end{picture}}}

\newcommand{\triplecrossed}{%
	\raisebox{\vertexheight}[2em][1.5em]{%
          \begin{picture}(70,25)(0,0)
		\put(5,0){\makebox(0,0){$\otimes$}} 
		\put(5,25){\makebox(0,0){$\otimes$}} 
		\put(35,0){\makebox(0,0){$\otimes$}} 
		\put(35,25){\makebox(0,0){$\otimes$}} 
		\put(65,0){\makebox(0,0){$\otimes$}} 
		\put(65,25){\makebox(0,0){$\otimes$}} 
	        \qbezier[20](9,4)(35,12.5)(61,21)
		\qbezier[20](9,21)(35,12.5)(61,4)
		\qbezier[8](35,5)(35,12.5)(35,20)		
		\linethickness{0.255em}
		\put(8.5,0){\line(1,0){22.5}}
		\put(8.5,25){\line(1,0){22.5}}
		\put(38.5,0){\line(1,0){22.5}}
		\put(38.5,25){\line(1,0){22.5}}
		\end{picture}}}

\newcommand{\vertex}[4]{
	\settowidth{\argwidth}{$#1$}
	\hspace{\argwidth}
	\raisebox{\vertexheight}[2em][1.5em]{%
        \begin{picture}(25,25)(0,0)
        	\put(0,0){\line(1,0){25}}
        	\put(0,25){\line(1,0){25}}
        	\put(8.5,0){\line(0,1){25}}
        	\put(16.5,0){\line(0,1){25}}
		\put(0,25){\hspace{-\argwidth}\raisebox{-0.5ex}{$#1$}}
		\put(25,25){\raisebox{-0.5ex}{$#2$}}
		\put(25,0){\raisebox{-0.5ex}{$#3$}}
		\settowidth{\argwidth}{$#4$}	
		\put(0,0){\hspace{-\argwidth}\raisebox{-0.5ex}{$#4$}}
   	\end{picture}}
	\settowidth{\argwidth}{$#3$}
	\hspace{\argwidth}
}
\newcommand{\clvertex}[4]{
	\settowidth{\argwidth}{$#1$}
	\hspace{\argwidth}
	\raisebox{\vertexheight}[2em][1.5em]{%
        \begin{picture}(25,25)(0,0)
        	\put(0,0){\line(1,0){25}}
        	\put(0,25){\line(1,0){25}}
	        \qbezier[15](12.5,0)(12.5,12.5)(12.5,25)
		\put(0,25){\hspace{-\argwidth}\raisebox{-0.5ex}{$#1$}}
		\put(25,25){\raisebox{-0.5ex}{$#2$}}
		\put(25,0){\raisebox{-0.5ex}{$#3$}}
		\settowidth{\argwidth}{$#4$}	
		\put(0,0){\hspace{-\argwidth}\raisebox{-0.5ex}{$#4$}}
   	\end{picture}}
	\settowidth{\argwidth}{$#3$}
	\hspace{\argwidth}
}

\newcommand{\twistedvertex}[4]{
	\settowidth{\argwidth}{$#1$}
	\hspace{\argwidth}
	\raisebox{\vertexheight}[2em][1.5em]{%
        \begin{picture}(25,25)(0,0)
        	\put(0,0){\line(1,0){25}}
        	\put(0,25){\line(1,0){25}}
	        \qbezier(8.5,0)(8.5,6.25)(12.5,12.5)
		\qbezier(12.5,12.5)(16.5,18.75)(16.5,25)
        	\qbezier(8.5,25)(8.5,18.75)(11.86,13.5)
		\qbezier(13.14,11.5)(16.5,6.25)(16.5,0)
 		\put(0,25){\hspace{-\argwidth}\raisebox{-0.5ex}{$#1$}}
		\put(25,25){\raisebox{-0.5ex}{$#2$}}
		\put(25,0){\raisebox{-0.5ex}{$#3$}}
		\settowidth{\argwidth}{$#4$}	
		\put(0,0){\hspace{-\argwidth}\raisebox{-0.5ex}{$#4$}}
   	\end{picture}}
	\settowidth{\argwidth}{$#3$}
	\hspace{\argwidth}
}

\newcommand{\PRL}{Phys.\ Rev.\ Lett.\ }
\newcommand{\JOB}{J.\ Opt.\ B:\ Quant.\ Semiclass.\ Opt.\ }
\begin{opening}
\title{Weak localisation of light by atoms \protect\\ with quantum internal structure}
\author{Cord A. M{\"u}ller}
\institute{Max-Planck-Institut f\"ur Physik komplexer Systeme\\
           N\"othnitzer Str.\ 38, D -- 01187 Dresden}
\author{Christian Miniatura}
\institute{Laboratoire Ondes et D\'esordre (FRE 2302 du CNRS)\\
	1361, route des Lucioles, F -- 06560 Valbonne} 

\end{opening}

\runningtitle{Weak localisation of light by atoms}

\begin{document}

\section{Introduction} 

\subsection{Why do we study quantum transport?}

As a commonplace, we could say that all simple problems in physics have
been 
solved a long time ago, and that we are tempted to turn to the 
challenging field called, rather pompously,  
``wave transport in complex systems''. Here, ``wave'' is ment to 
emphasize the influence of \emph{interference} (which is truly
quantum mechanical for massive particles like electrons or atoms). 
``Transport'' implies that we are interested in situations \emph{out of
thermodynamic equilibrium} (but not too far, so that linear response theory
applies). Finally, a system will be called ``complex'' whenever it is
\emph{disordered}, or 
\emph{strongly interacting}, or \emph{chaotic}. 
Our choice of this field is motivated by two aspects:   
on the side of applied physics, all remote sensing techniques 
need to incorporate the multiple scattering of waves in turbid
media, and the miniaturisation process in semi-conductor industry arrives
at length scales where the control of quantum interference becomes crucial.  
On the academic side of more fundamental physics, we would like to 
understand and enjoy the predictive power  of the best
physical theory available today: quantum theory.  

The classical paradigm of transport in a disordered environment is
\emph{diffusion}.  In general, a quantity $n(x,t)$ in space and time 
(think of a particle concentration) at equilibrium is distributed with a
constant 
density $n_0$. If the quantity is locally conserved, 
a small variation $\delta n(x,t)$ then obeys the diffusion equation, 
$(\partial_t-D_0\nabla^2)\delta n(x,t)=0$; $D_0$ is the diffusion constant. 
Its solution in Fourier space 
\eqlab{
\delta \hat n(q,t)=\delta \hat n(q,0) \exp(-D_0q^2t)
}{diffusion}
shows that very smooth fluctuations in real space (with small wave numbers
$q \to 0$) are removed on very long time scales $1/D_0q^2\to\infty$. Indeed,  
because of the local
conservation law, a small surplus $\delta n>0$ cannot just simply disappear,
but has to reach a place far away 
with a depleted density $\delta n<0$ in order to restore
the equilibrium situation $\delta n=0$.  

Classical Boltzmann diffusion theory has
been used successfully to describe electric conduction 
in metals (first by Drude),
or the diffusion of light intensity through stellar atmospheres or 
interstellar clouds
(notably by Schuster and Chandrasekhar)~\cite{cldiffusion}. 
However, at the microscopic level, one has to deal with waves. As an 
example, the quantum picture of
electrons moving in a perfectly periodic crystal pinpoints the key role 
of interference: an electron, 
initially confined in a well, can resonantly tunnel through the lattice,
yielding Bloch energy bands.   
In trying to 
understand the interplay between disorder and interference, 
P.\ Anderson showed that  
sufficiently strong disorder can suppress the quantum diffusion (leading to a
vanishing diffusion constant $D=0$), a
phenomenom baptised \emph{localisation}~\cite{Anderson58}. 
Later, it was realized that
even far from the regime of (strong) localisation, diffusive transport is
affected by interference: this so-called 
\emph{weak localisation} 
reduces the Boltzmann diffusion constant,  $D=D_0-\delta D$, by the
constructive pairwise interference of amplitudes associated with time-reversed
scattering paths. 
Mesoscopic physics, namely the study of interference 
effects in wave transport through random media, was born \cite{Houches94}.

\subsection{Why do we use photons and atoms?}

The theory of localisation has been first 
developed in the condensed matter physics
community. But as electrons are charged particles with a very strong and
long-range Coulomb interaction, the pure one-particle picture of Anderson
localisation has never been observed experimentally as far as we know. 
On the other hand, the
radar physics community first 
had realized that interference of counter-propagating
amplitudes can play an important role in the multiple scattering of
electromagnetic waves~\cite{Watson69}. Taking the better of the two worlds, 
S.\ John and P.\ Anderson suggested to study localisation using light or other
non-interacting classical waves. Light scattering 
allows the use of modern 
lasers with excellent coherence properties as well as an accurate analysis of
direction and polarisation. A large number of turbid media has thus
been studied, from Saturn's rings to semi-conductor powders~\cite{CBS}.  

Today, laser-cooled atoms as light scatterers come close to 
a theoretician's dream team: they are perfectly
identical (monodisperse) point particles. But more than that (theoretical
convenience hardly ever justifies expensive experiments): their specific
properties permit to 
study new regimes which are characterised, for example, by their quantum
internal structure (as discussed in the present contribution),  
the finite width of the atomic fluorescence
spectrum, non-linearities such as the saturation of an atomic transition, and
the  
mechanical acceleration of the atom due to light scattering. 
Even more, at sufficiently low temperature, the atoms themselves become matter
waves and can in turn be used to probe quantum transport, all the way
down to the degenerate quantum regime of Bose-Einstein condensation (see, for
example, the recent 
beautiful experimental realisation of the Mott-Hubbard quantum
phase transition by the Munich BEC group \cite{Greiner02}).

\subsection{Why do we need to describe an internal structure?} 

\begin{figure}
\begin{center}
\psfrag{a}{$\alpha$}
\psfrag{Polystyrene}{Styrofoam}
\psfrag{Rb 85}{cold Rb$^{85}$}
\psfrag{theta}{$\theta$ (mrad)}
\psfrag{hpar}{$\hpar$}
\psfrag{hperp}{$\hperp$}
\includegraphics[width=\textwidth]{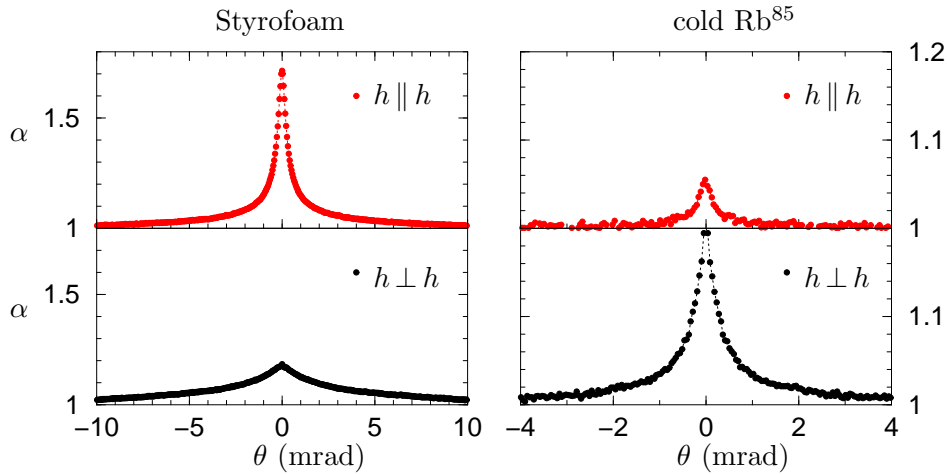}
\caption{Experimental CBS enhancement $\alpha$ vs. the angle $\theta$ with
respect to the backscattering direction
for a classical medium (Styrofoam, left) and cold rubidium atoms (right), in
the channel of preserved helicity ($\hpar$, top) and flipped helicity
($\hperp$, bottom). Data generously provided by G.\ Labeyrie.}
\label{expcone.fig}
\end{center}
\end{figure}

The first experiment of coherent backscattering of light from a cloud of cold
atoms in 1999 \cite{Labeyrie99,Labeyrie00}
yielded a surprising result (see \refig{expcone}): the observed interference
peak in the channel of preserved helicity ($\hpar$) shows only a maximum
enhancement of about 1.05, far below the classically expected
factor of 2.0 due to reciprocity \cite{Maynard}. Also, this $\hpar$ enhancement
is much smaller than the measured value of 1.2 in the channel of flipped
helicity ($\hperp$). It was soon realized that the
\emph{degeneracy} of the probed atomic 
dipole transition $J=3 \to \Je=4$ is responsible for an 
\emph{imbalance of CBS amplitudes} 
and therefore reduces the measurable enhancement factor 
\cite{Jonckheere00}. We thus have to generalise 
the theory of the multiple scattering of
polarised light by point dipoles to the case of an arbitrary atomic transition
$J \to \Je$~\cite{CMthesis}. This theory indeed explains the observed
enhancement factors 
and shall be described in the following.   

\section{Multiple scattering of a photon by atoms with internal degeneracy}

\subsection{The one-photon transition matrix} 

\begin{figure}
\begin{center}
\psfrag{Gamma}{$\Gamma$}
\psfrag{delta}{$\delta{\tiny\{}$}
\psfrag{omega0}{$\omega_0$}
\psfrag{Jeme}{$\ket{\Je\me}$}
\psfrag{Jm}{$\ket{Jm}$}
\includegraphics[width=\textwidth]{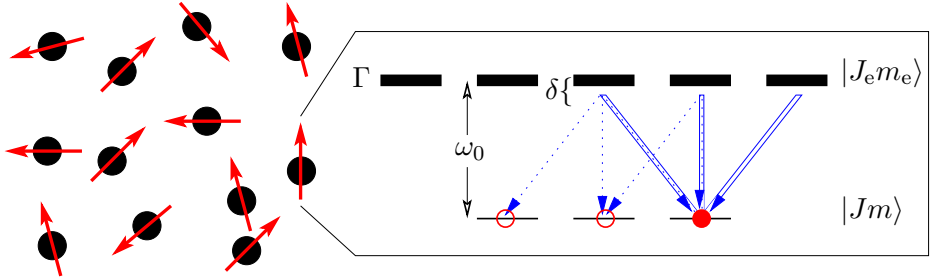}
\caption{Left: a cloud of 
cold atoms as point scatterers with fixed random positions $\bs r$ 
and randomly
oriented total angular momenta $J$.  
Right: a zoom into the energy level scheme of a resonant degenerate dipole
transition, here 
the example of $J=1$, 
$\Je=2$. $\delta=\omega-\omega_0$ is the detuning of
the probe light from the atomic resonance frequency $\omega_0$, and  $\Gamma$
is the natural width of the
excited atomic state. The polarisation of scattered polarised photons (full and
dotted lines) is coupled to the internal magnetic quantum numbers $m$.}
\label{samplezoom.fig}
\end{center}
\end{figure}

Consider a cloud of laser-cooled atoms confined in a standard magneto-optical
trap. The cooling is such that 
their velocity spread $v$ is much smaller than the Doppler velocity $\Gamma/k$
($\Gamma$ is the natural width of the excited atomic state).
Therefore, we can neglect the Doppler effect and may 
assume that the atoms' positions $\bs
r_\alpha,\alpha=1,\dots,N$, remain fixed on the 
light-scattering time scale.  
On the other hand, the velocity spread should be 
much larger than the recoil velocity
$v_{\text{rec}}=\hbar k/M$ (where $M$ is the atomic mass) 
for the scattering of a photon of wave-vector $k$. This allows us to treat the
positions as classical random variables and to follow the standard
diagrammatic approach to describe  multiple scattering (see \cite{vanRossum99}
and references therein).  
The CBS probe beam with incident wave-vector $\bs k$, polarisation $\beps$ and
frequency $\omega$ excites a closed atomic dipole transition
defined by a 
ground state with total angular momentum $J$ and an excited state $\Je$ 
with frequency $\omega_0$. In the absence of a magnetic field, these 
two levels with internal quantum numbers $m$ and $\me$ 
are respectively $(2J+1)$- and $(2\Je+1)$-fold degenerate (see
\refig{samplezoom}). 

In order to minimise saturation effects and optical pumping, the experiment is
performed at low laser intensity, so that we can consider the light scattering
in the limit of one-photon Fock states $\ket{\bs k\beps}$ (in this notation,
the transversality $\beps\cdot\bs k=0$ is understood). 
The transition amplitude for the scattering of an
incident photon $\ket{\bs k\beps}$ into an emitted photon $\ket{\bs k'\beps'}$
by a single atom is 
\eqlab{%
\raisebox{-12mm}[20pt][15pt]{\includegraphics{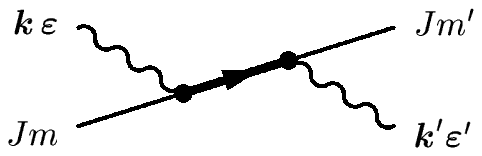}} = \bra{Jm'}\bbeps'\cdot
\bs t(\omega)\cdot\beps\ket{Jm}\,e^{i(\bs k-\bs k')\cdot \bs r}.
}{singamp} 
Note that the exponential with the classical external degrees of freedom $\bs
r$ is factorised from the matrix element with the internal quantum numbers.  
Therefore, the usual multiple scattering formalism applies~\cite{vanRossum99}. But  we
have to analyse carefully the role of the internal degrees of freedom. 
We see in \refeq{singamp} that the incident and emitted
polarisation vectors $\beps$ and $\bbeps'$ (the bar denotes complex
conjugation) are coupled by the transition matrix
$t_{ij}(m,m';\omega)=\bra{Jm}t_{ij}(\omega)\ket{Jm}$.  
As any $3\times 3$ matrix,
it can be decomposed into its irreducible components with respect to
rotations, 
\eq{t_{ij}= \underbrace{\frac{1}{3}\delta_{ij}t_{kk}}_{t^{(0)}_{ij}} 
	+ \underbrace{\frac{1}{2}(t_{ij}-t_{ji})}_{t^{(1)}_{ij}}  
+ \underbrace{\frac{1}{2}(t_{ij}+t_{ji})-\frac{1}{3}\delta_{ij}t_{kk}}_{t^{(2)}_{ij}}}  
its scalar part or trace $t^{(0)}$, its antisymmetric part
$t^{(1)}$ and its traceless symmetric part  
$t^{(2)}$
(summation over repeated indices is understood). It is easy to verify that in
the non-degenerate limit $J=0$, corresponding to the classical case of an
isotropic dipole, only the scalar component $t^{(0)}_{ij}$ exists. So our task
in the following will be to determine how the non-scalar
components $t^{(1,2)}_{ij}$ influence the light propagation.

\subsection{Average propagation inside the effective medium}

In a disordered environment, the actual propagation depends on the precise
realisation of disorder and yields, for example, an interference
``finger-print'' or speckle-pattern. 
Universal properties can only be obtained on average. The
average propagation of a light mode $\ket{\bs k\beps}$ inside the medium
is described by the average propagator $\mv{G_{ij}(\bs k,\omega)}$
where the average $\mv{\dots}$ traces out the matter degrees of
freedom.  
Following the standard procedure, we determine the average propagator 
in terms of the vacuum
propagator $g_0$ and the self-energy $\Sigma$ with the aid of the Dyson
equation (tensor indices are omitted for brevity)  
\eqlab{\mv{G}= g_0 + g_0 \Sigma \mv{G}= \frac{1}{g_0^{-1}-\Sigma}.
}{dyson}  
For a sufficiently dilute scattering medium such that $n\lambda^3\ll 1$ ($n$
is the atomic number density with typical values of $6\times10^{10}\,$
cm$^{-3}$, and $\lambda$ is the optical wavelength, typically $780\,$nm in the
experiments done so far),
the independent scattering approximation applies, and the self-energy is
simply the average single-scattering t-matrix:
$\Sigma_{ij}(\omega)=N\mv{t_{ij}(\omega)}$. 
We suppose that the atomic scatterers are \emph{distributed
uniformly} over their internal states such that rotational invariance is
restored on average: $\Sigma_{ij}(\omega) = \delta_{ij}\Sigma(\omega)$. 
This is intuitively clear since 
under a scalar average only the scalar component can survive. 
Now, it can be checked that
the scalar component is precisely the one that mimics a classical isotropic
dipole:
\eq{ \Sigma(\omega)= n M_J\frac{3\pi}{k^2} \frac{\Gamma/2}{\delta+i\Gamma/2}.
}
Here, the quantum internal structure only enters through the scalar factor
$M_J=(2\Je+1)/3(2J+1)$ with the non-degenerate limit $M_0=1$. 
The elastic mean free time is defined as $\tau=-(2\Im\Sigma(\omega))^{-1}$ 
and yields the mean free path $\ell=\tau$ in our units, where $c \equiv 1$. 
By virtue of 
the optical theorem, the mean free path is related to the 
total cross-section for elastic scattering by the usual Boltzmann
expression $\ell=1/n\sigma$. 
 
Obviously, the quantum internal structure (or contribution of 
non-scalar components of the t-matrix) has 
disappeared under the scalar average over internal states, and
we recover a standard scalar theory. Have we been too optimistic in hoping
that the internal structure can explain the dramatic reduction of interference
in the CBS signal? Well, of course not: the experimental signal in
\refig{expcone} is the
\emph{average intensity} which must be distinguished from the square of the
average amplitude. More technically speaking, after calculating the average
amplitude $\mv{G}$ with the Dyson equation \refeq{dyson}, we have to turn to
the average intensity $\mv{\overline{G}G}$.

\subsection{The average intensity vertex}

In a dilute medium $n\lambda^3\ll 1$, the building block for the multiple
scattering series  
is the average single scattering intensity vertex $\mv{t_{ij}\bar
t_{kl}}$~\cite{Lagendijk96}. 
It connects two amplitudes with their respective polarisation
vectors and can thus be written as a rank-four tensor, 
\eq{\mv{t_{ij}(\omega)\bar t_{kl}(\omega)} = u(\omega) \vertex{i}{j}{k}{l}.
} 
Here, the four point polarisation vertex can be calculated analytically using
standard methods of irreducible tensor operators~\cite{Mueller01}. It is
simply given as the 
sum of the three pairwise contractions, 
\eqlab{\vertex{i}{j}{k}{l}=w_1 \delta_{ij}\delta_{kl}
	+w_2 \delta_{ik}\delta_{jl}
	+w_3 \delta_{il}\delta_{jk}
}{defvertex} 
where the coefficients $w_1=\frac{1}{3}(s_0-s_2)$,
$w_2=\frac{1}{2}(s_2-s_1)$ and $w_3=\frac{1}{2}(s_2+s_1)$ 
are given in terms of the squared $6J$-symbols
\eq{%
	s_K=3(2\Je+1)\sixj{1}{1}{K}{J}{J}{\Je}^2
}
associated with the irreducible t-matrix components $t^{(K)}$ of 
order $K=0,1,2$. 
For the classical isotropic dipole $J=0$, the coefficients become
$(w_1,w_2,w_3)=(1,0,0)$ and yield the purely horizontal contraction
$\delta_{ij}\delta_{kl}$. 
We see therefore that including the quantum internal
structure is equivalent to replacing the simple dotted line in the
usual diagrams~\cite{vanRossum99}  
by the generalised vertex \refeq{defvertex} with the richer 
topology of a \emph{ribbon}:
\eqlab{\clvertex{i}{j}{k}{l}\ \stackrel{J>0}{\longrightarrow}\ \vertex{i}{j}{k}{l}
}{genvertex}   

\subsection{Summation of ladder and crossed series}

Following the standard diagrammatic approach, we have to sum the so-called
series of ladder diagrams
\eqlab{%
\intensbox{L} = \lstep + \lstep \avgintens\lstep 
+ \lstep \avgintens \lstep \avgintens \lstep + \dots 
}{ladderseries}
that describe the average intensity neglecting all interference terms: the
direct amplitude (upper line) is scattered by the same scatterers as the
conjugate amplitude (lower line), as indicated by the dotted lines. At least
formally, one recognises a geometrical series that may be summed up
analytically once the single scattering vertex (first term on the r.h.s.) and
the square of the average propagators (thick lines) are known. The
interference correction associated with CBS and weak localisation is contained
in the so-called maximally crossed diagrams 
 \eqlab{
\intensbox{C} =  \crossed + \triplecrossed +\dots 
}{crossedseries}  
that describe the propagation of the direct and conjugate amplitude along the
same scattering path, but in opposite directions. For classical point
scatterers, the crossed and ladder diagrams are closely related by 
reciprocity: by turning around the lower line of a maximally crossed diagram,
the connecting lines straighten out and yield the corresponding ladder
diagram. By carefully counting incident and emitted momenta and polarisation
indices, one shows that the diagrams are rigorously equal for scattering in
the backward direction ($\bs k'=- \bs k$) and in the parallel polarisation
channels $\bbeps'=\beps$. This identity justifies the CBS enhancement factor of
2.0 in the helicity-preserving polarisation channel where, for 
spherically-symmetric scatterers, the single scattering contribution 
is absent.

In our case of a quantum internal structure, see \refeq{genvertex}, 
the classical vertex has to 
be substituted by the ribbon vertex.  But now the
correspondence between ladder and crossed diagrams is spoilt: when the bottom
line of a crossed diagram is turned around, the connecting ribbons are
twisted:
\eq{\vertex{}{}{}{}\ \rightarrow\ \twistedvertex{}{}{}{}\ \ne \vertex{}{}{}{}\
.}
The evident difference between the twisted and the straight ribbon shows that
the quantum internal structure of the atomic scatterer indeed affects the
interference corrections to the average intensity. More quantitatively,
twisting the vertex is equivalent to the exchange of the coefficients $w_2$
and $w_3$ associated with the diagonal and vertical contractions. Once we have
calculated the ladder series, we can then obtain the crossed series by simply
replacing $w_2\leftrightarrow w_3$ (and rearranging the momenta and
polarisation vectors as in the classical case).

In order to sum the geometrical ladder series, we have to determine the
eigenvalues of the atomic vertex with respect to the ``horizontal'' direction
of summation. Using a basis of projectors $\sfT^{(K)}$ onto irreducible
eigenmodes 
$[\eps_i\bar\eps_l]^{(K)}$ of the field polarisation matrix, we obtain a
decomposition of the form
\eq{%
\vertex{i}{j}{k}{l}=\sum_K\lambda_K(J,\Je)\sfT^{(K)}_{il,jk}
}      
where the eigenvalues $\lambda_K$ are simple functions of the $w_i$:
\eq{\lambda_0=1,\qquad\lambda_1=w_1-w_2,\qquad \lambda_2=w_1+w_2.}
Taking
advantage of our simple substitution rule $w_2\leftrightarrow w_3$, the
twisted vertex of the crossed series then gives crossed eigenvalues $\chi_K$
such that  
\eq{ %
\twistedvertex{i}{j}{k}{l}=\sum_K\chi_K(J,\Je)\sfT^{(K)}_{il,jk}. 
} 
Having diagonalised the scattering vertex, we have to treat also the transverse
propagation between atoms. The actual calculation is rather cumbersome because
it involves momentum-dependent eigenvalues and
projectors; details can be found in \cite{Mueller02}. 
But our approach permits us
to sum analytically the ladder and crossed series, for the full transverse
vector field with arbitrary polarisation, and for arbitrary atomic
transitions. It is a first step towards a generalisation of the existing
multiple scattering theories, either of scalar waves by anisotropic
point-scatterers~\cite{Ozrin92b,Amic96}, or of vector waves by isotropic
Rayleigh scatterers~\cite{Ozrin92a,Amic97}.

\section{Bulk transport properties} 

\subsection{Diffusion and depolarisation}
\label{depolar.sec}

The summed ladder propagator describes the average intensity distribution
inside the bulk medium (\emph{i.e.} in the absence of boundaries). To gain
qualitative insight, we can simplify the exact expressions by the diffusion
approximation (retaining terms of order $q^2$). The crucial ingredients are
the atomic eigenvalues $\lambda_K$ and $\chi_K$ as well as the
eigenvalues $b_0=1$, $b_1=\frac{1}{2}$ and $b_2=\frac{7}{10}$ 
associated with the transverse propagation.
The ladder propagator in the
long-time limit then takes the form 
\eq{%
L(q,t)=\sum_K l_K 
	\exp\left[ -D_Kq^2t-t/\tau_{\text{p}}(K)\right].
} 
Apart from the factor $l_K$ with no importance here,  
we first notice an exponential decay that was anticipated above in the general
diffusion picture 
\refeq{diffusion}. We obtain the diffusion constant $D_K\approx D_0=\ell
v_{\text{tr}}/3$ in terms of the well-known transport velocity
$v_{\text{tr}}=\ell/\tau_{\text{tr}}$ of resonant point scatterers~\cite{vanRossum99}. But there is
a second
exponential decay, with a characteristic polarisation relaxation time
\eqlab{%
\tau_{\text{p}}(K)=\frac{\tau_{\text{tr}}}{1/b_K\lambda_K-1}. 
}{ladder}
The scalar field mode $(K=0)$ describes the total intensity. Its relaxation
time is infinite $(\tau_{\text{p}}(0)=\infty)$ for all atomic transitions
(since $b_0=\lambda_0=1$), 
and we recover a purely diffusive behaviour as required by
energy conservation. The non-scalar field modes are exponentially damped on 
finite, and in fact rather short time scales $\tau_{\text{p}}(1,2)\le
\tau_{\text{tr}}$: this simply says that a well-defined field polarisation is
scrambled in the course of multiple scattering. 
With increasing degeneracy $J>0$ of the atomic transition,
these 
times get even shorter ($\lambda_K<1$), which nicely confirms the intuitive
picture that random transitions between different atomic Zeeman substates
enhance the depolarisation.   

\begin{figure}
\begin{center}
\fbox{\small
\psfrag{e}{$\beps$}
\psfrag{e'}{$\beps'$}
\psfrag{e''}{$\beps''$}
\includegraphics[width=0.55\textwidth]{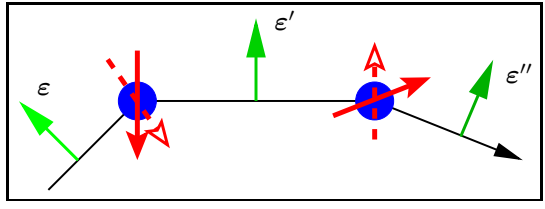}}
\caption{The transversality of propagation depolarises the multiply scattered
light. This depolarisation is enhanced by an atomic quantum internal structure
$J>0$ (as discussed in sec.~\ref{depolar.sec}). Furthermore, the internal
degeneracy 
leads to dephasing of the interference correction (as discussed in sec.~\ref{dephase.sec}).}
\label{}
\end{center}
 \end{figure}

\subsection{Weak localisation and strong dephasing}
\label{dephase.sec}

Under the same approximations, the sum of crossed diagrams yields a
contribution 
\eq{%
C(q_c,t)=\sum_K c_K \exp\left[-D_Kq_c^2t -t/\tau_{\text{p}}(K)-t/\tau_\phi(K)\right] 
} 
as a function of the total momentum $q_c=||\bs k+\bs k'||$. 
One first recognises the same exponential
decay as in the ladder contribution \refeq{ladder}, indicating that
depolarisation also 
affects the coherent contribution. But there is an additional source of
exponential damping described by dephasing times~\cite{Akkermans02}
\eqlab{
\tau_\phi(K)=\frac{b_K\tau_{\text{tr}}}{1/\chi_K-1/\lambda_K}
}{tauphi}
which we define precisely as the damping times \emph{with respect to} the
incoherent depolarisation times. 
Of particular interest is the dephasing time
$\tau_\phi(0)$ of the scalar mode or intensity. For atomic transitions of the
type $\Je=J+1$, we find explicitly
\eq{
\tau_\phi(0)=\frac{\tau_{\text{tr}}}{J(2J+3)}. 
}  
The interference is only preserved ($\tau_\phi=\infty$) for classical dipoles
$J=0$. For the least possible degeneracy $J=\frac{1}{2}$, the dephasing time
is already as short as $\tau_{\text{tr}}/2$ and decreases as $1/J^2$. Even if
the exact expression becomes meaningless for larger $J$ 
(since the diffusion approximation is
certainly not justified on time scales shorter than the transport time), it is
evident that the internal structure destroys the interference very
effectively. Therefore, in the presence of a quantum internal structure, we
only expect a (very, very) weak localisation correction to the Boltzmann
diffusion constant of light propagation.

\section{Coherent backscattering}

We wish to calculate 
the CBS peak~\cite{CBSpeak} 
analytically for arbitrary atomic transitions and therefore 
choose the simplest possible geometry of the scattering medium, a
semi-infinite half-space.  
Having calculated the bulk propagator $F_0(\bs r_1-\bs r_2)$ (either the
ladder or the crossed contribution), 
we define the corresponding propagator for the
semi-infinite half space by using the method of images,  
$	F(\bs r_1,\bs r_2)=F_0(\bs r_1-\bs r_2)-F_0(\bs r_1-\bs r_{2'})$. 
The image point $2'$ is defined with respect to a
mirror plane lying at a 
distance $z_0$ outside the boundary of the medium. Following the habits of the
field, we use the so-called ``skin
layer depth'' $z_0 \approx 0.7121 \, \ell$ pertaining to the exact
solution of the homogeneous Milne equation for vector waves and point dipole
scatterers~\cite{Ozrin92a,Amic97}.   

\begin{figure}
\begin{center}
\psfrag{J}{\makebox[0pt]{$J=\Je-1$}}
\psfrag{hpar}{$\hpar$}
\psfrag{hperp}{$\hperp$}
\psfrag{lpar}{$\lpar$}
\psfrag{lperp}{$\lperp$}
\psfrag{a}{\makebox[0pt]{$\alpha$}}
\psfrag{Experiment}{Rb experiment}
\psfrag{Polarization channel}{Polarisation:}
\includegraphics[width=0.75\textwidth]{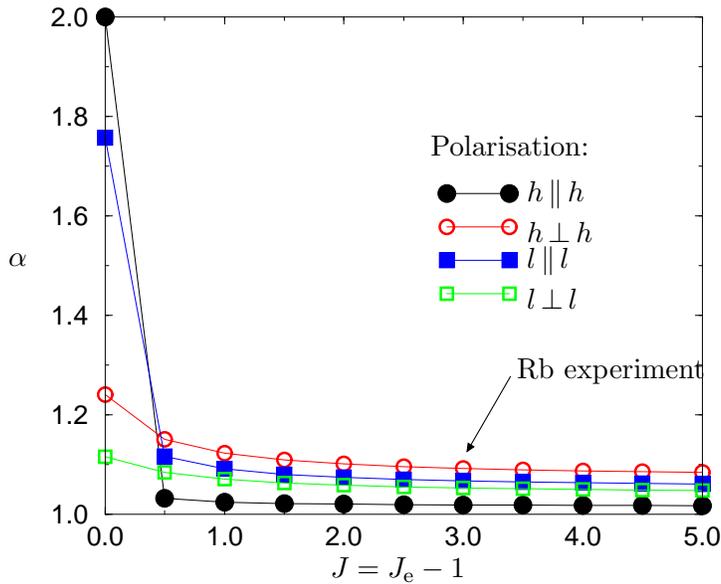}
\caption{Calculated CBS enhancement factors at exact backscattering from a
homogeneous 
half-space of atoms with a degenerate dipole transition $J\to J+1$ for the four
usual channels of preserved ($\parallel$) and flipped ($\perp$)
linear polarisation ($l$) or helicity ($h$). Contrary to the classical case,
for atoms with $J>0$ the best enhancement is expected in the $\hperp$ channel.}
\label{ampli.fig} 
\end{center}
\end{figure}

By this approach, we can calculate the CBS enhancement factors and peak shapes
\emph{beyond the diffusion approximation} (which is crucial whenever only
short paths contribute). The CBS enhancement
factor $\alpha$ is plotted in \refig{ampli} as a function of $J$ for transitions of the type $J\to
J+1$. In the classical limit $J=0$, we recover values of $\alpha=2.00$,
$1.76$, $1.24$ and $1.12$ for the different polarisation channels which are in
excellent agreement with the exact values obtained by a solution of the vector
Milne equation by the Wiener-Hopf method~\cite{Ozrin92a,Amic97}. 
This indicates that the method of images can be used with success even for
signals involving short scattering paths (like in the perpendicular channels)
provided that the \emph{exact} propagator (beyond the diffusion approximation)
be used.

Figure \ref{ampli.fig} shows that the least internal degeneracy reduces the
CBS interference dramatically: the perfect factor of $2.0$ in the $\hpar$
channel plunges down to $1.04$, well below the other three polarisation
channels. At this point, our theory 
indeed explains the astonishing experimental
result shown in \refig{expcone} that has motivated this work.  Furthermore,   an  experiment using cold Strontium atoms without internal degeneracy
($J=0$) has recently 
confirmed that excellent enhancement factors in agreement with the
theoretical predictions are obtained ~\cite{Bidel02}. 
 
\begin{figure}
\begin{center}
\psfrag{a}{$\alpha$}
\psfrag{mu}{$k \ell\theta $}
\psfrag{hpar}{$\hpar$}
\psfrag{hperp}{$\hperp$}
\psfrag{lpar}{$\lpar$}
\psfrag{lperp}{$\lperp$}
\psfrag{j=0,je=1}{\makebox[0.75cm]{$\Jg=0$, $\Je=1$:}}
\psfrag{j=3,je=4}{\makebox[0.75cm]{$\Jg=3$, $\Je=4$:}}
\includegraphics[width=\textwidth]{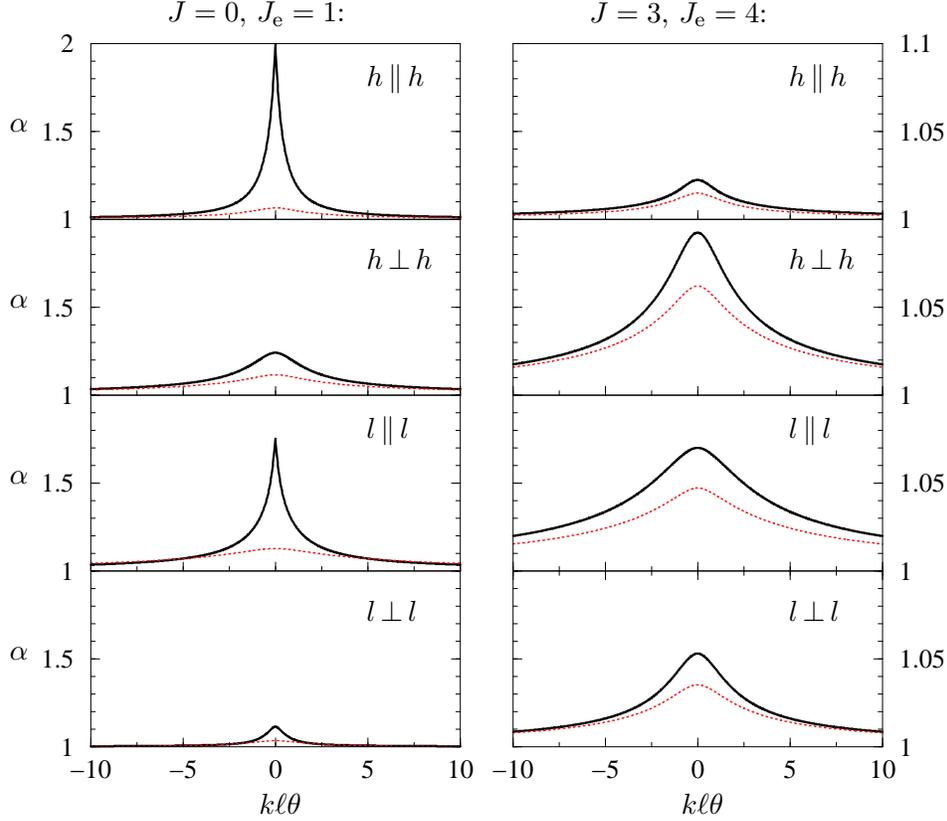}
\end{center}
\caption{Calculated CBS enhancement factor $\alpha$ as a function of the
reduced 
backscattering angle $k \ell \theta$ for isotropic dipole scatterer (left) and
atoms with a degenerate $J=3\to \Je=4$ transition (right). Full line: sum of
all scattering orders. Dotted line: double
scattering contribution.}
\label{imcone.fig}
\end{figure}

Figure \ref{imcone.fig} shows a comparison between calculated CBS peak shapes for isotropic point scatterers
(left side) and atoms with $J=3\to \Je=4$ (right side). The full lines
are the sum of all scattering orders, and the dotted lines indicate the double
scattering contribution (which is known in closed form~\cite{Mueller01}). 
Whereas the CBS peak in the classical case and
parallel polarisation channels contains very high orders of scattering
(corresponding to long scattering paths), the atomic internal degeneracy cuts
off these contributions (as indicated already by the dephasing times
\refeq{tauphi}) and yields only very small peaks.   

A quantitative comparison of the theory to the experimental results needs to
take into account the finite geometry of the actual atomic cloud (roughly
spherical, with a Gaussian density distribution). This means that analytical
results are out of reach, and a numerical approach has to be taken. A Monte
Carlo simulation of photon trajectories in various geometries has been
realized by D.~Delande and yields results which are in good agreement with the
experimental data~\cite{Labeyrie02}.

\section{Concluding remarks and acknowledgements}

In summary, 
we have presented an analytical theory of the multiple scattering of polarised
photons by resonant atomic dipole transitions with arbitrary degeneracy. We
have shown how the usual diagrammatic approach can be generalised by using an
intensity vertex with a ``ribbon'' topology that breaks the equivalence
 of ladder and crossed diagrams. 
The theoretical CBS peak heights reproduce the experimental
results: the quantum internal structure indeed reduces the CBS interference
drastically. 
Inasmuch as weak localisation acts as a precursor for Anderson localisation,  
our results indicate that in order to reach the strong localisation
regime with cold atoms, the use of a non-degenerate transition is highly
recommendable.  

We hope to have demonstrated that 
atoms are light scatterers with intriguing interference properties. They permit
to set up a microscopic theory for diffusion and weak localisation and thus
promise to be a source of further inspiration for both fields of ``disordered
systems/multiple scattering'' and ``atomic physics/quantum optics''.

It is a pleasure to acknowledge the invaluable help by Dominique Delande
(who is to be credited for the ``exact-image'' method), many inspiring
discussions with Eric Akkermans (who is to be credited for the ``dephasing''
interpretation), and a critical reading of the manuscript by Andreas
Buchleitner.

\end{document}